\documentclass[12pt,preprint]{aastex}

\def\hMsun{h^{-1} \; M_\odot}
\def\hMpc{h^{-1} \; {\rm Mpc}}

\newcommand{\appropto}{\mathrel{\vcenter{
  \offinterlineskip\halign{\hfil$##$\cr
    \propto\cr\noalign{\kern2pt}\sim\cr\noalign{\kern-2pt}}}}}
\usepackage{pslatex}
\usepackage{graphicx}
\begin{document}

\title{Halo Occupation Distribution of Infrared Selected Quasars} 
\author{Kaustav Mitra UG-3, \\Registration Number: 13120911012\\Department of Physics, Presidency University\\ 86/1 College Street, Kolkata 700073\\
PHYS-0692\\ Advisor: Suchetana Chatterjee}


\section*{Abstract}
We perform a Halo Occupation Distribution (HOD) modeling of the projected two-point correlation function (2PCF) of quasars  that are observed in the Wide-field Infrared Survey Explorer (WISE) telescope with counter-parts in the Sloan Digital Sky Survey (SDSS) Data Release (DR)-8 quasar catalog at a median redshift of $z\sim 1.04 (\pm 0.58)$. Using a four parameter HOD model we derive the host mass scales of WISE selected quasars. Our results show that the median halo masses of central and satellite quasars lie in the range $M_{\mathrm{cen}} = (5 \pm 1.0) \times 10^{12} M_{\odot}$ and $M_{\mathrm{sat}} = 8 (^{+7.8} _{-4.8}) \times 10^{13} M_{\odot}$, respectively. The derived satellite fraction is $f_{\mathrm{sat}}= 5.5 (^{+35} _{-5.0})\times 10^{-3}$. Previously Richardson et al.\ used the SDSS DR7 quasar clustering data to obtain the halo mass distributions of $z\sim 1.4$ quasars. Our results on the HOD of central quasars are in excellent agreement with Richardson et al.\ but the host mass scale of satellite quasars for the WISE sample, is lower than that of Richardson et al.\ resulting in an order of magnitude higher satellite fraction for the WISE sample. We note that our sample of quasars are systematically brighter in the WISE frequency bands compared to the full quasar sample of SDSS. We discuss the implication of this result in the context of current theories of galaxy evolution.


\clearpage
\tableofcontents
\clearpage
\section{Introduction}

According to the cold dark matter paradigm of galaxy formation, it is believed that galaxies form in the potential wells of giant dark matter (DM) halos \citep[e.g.,][]{w&r78, w&f91, kauffmannetal93, nfw95, m&w96, kauffmannetal99} .Observationally it is now well known that at the center of every massive galaxy in the Universe lies a supermassive black hole \citep[(SMBH) e.g.,][]{soltan82, tremaineetal02}. The bolometric emission from the central SMBH in some galaxies outshines the emission from the entire galaxy. These classes of galaxies are called active galactic nuclei (AGN). It has been also established that galaxy evolution and growth of supermassive black holes are intrinsically linked \citep[e.g.,][]{m&f01, tremaineetal02, grahametal11}. So to study different stages of galaxy evolution, we would need to understand the connection between the growth and formation of SMBH along with their host galaxies and dark matter halos (known as AGN/SMBH co-evolution in the literature).

The co-evolution of SMBH with DM halos has been studied via analytic techniques and numerical simulations \citep[e.g.,][]{k&h00, w&l03, marconietal04, cattaneoetal06, crotonetal06, hopkinsetal06, lapietal06, shankaretal04, dimatteoetal08, b&s09,  volonterietal11, c&w13}. The key observational probe to understand the relation between SMBH and their host halos has been through the measurement of the two-point-correlation function \citep[2PCF; e.g.,][]{arp70}. Clustering measurements of different types of AGN have been carried out by several groups employing data from multiple surveys \citep[e.g.,][]{myersetal06, myersetal07a, coiletal07, shenetal07, wakeetal08, shenetal09, rossetal09, coiletal09, hickoxetal11, allevatoetal11, whiteetal12, shenetal12a, krumpeetal12, mountrichasetal13, koutoulidisetal13}.

Majority of these studies involve clustering measurements of a certain class of AGN, namely optically bright quasars. Due to their high luminosity, quasars are detected to $z \gtrsim 7$ \citep[e.g.,][]{mortlocketal11}, making them powerful probes of structure formation over a broad range of redshifts. In addition, the large sample sizes of quasars and the availability of spectroscopic redshifts make them excellent candidates for studying AGN co-evolution with cosmic structures. However, quasars have broad spectral-energy distributions and the emission at different wavelengths are sensitive to different physical processes within the central SMBH. Recently, lot of efforts have been put forward to study the clustering properties of quasars that have been selected in other wavebands too \citep[e.g.,][]{shenetal09, donosoetal10, hickoxetal11, donosoetal14, dipompeoetal14, dipompeoetal16}. In this work, we use the clustering data from the  Wide-field Infrared Survey Explorer (WISE) selected quasars (with SDSS counterparts) from \citet{dipompeoetal14} (D14 hereafter) and for the first time employ the halo occupation distribution (HOD) formalism \citep[e.g.,][] {m&f00, seljak00, b&w02, zhengetal05, z&w07, wakeetal08, shenetal10, miyajietal11, starikovaetal11, allevatoetal11, richardsonetal12, k&o12, shenetal12a, richardsonetal13} to derive their host dark matter halo properties. 

\citet{richardsonetal12} (R12 hereafter) performed a similar HOD modeling of the 2PCF of SDSS-DR7 quasars at a median redshift of $z\sim 1.4$, which is similar to the typical redshifts of D14 quasars. We compare the HOD properties of the infrared selected quasars with the optically bright sample and show that although the large scale environments of these two classes of quasars are similar, there exist significant differences in the small scale environments. We then discuss the implication of this result in the context of theories of galaxy evolution and the associated quasar activities in galaxies. 

The report is organized as follows. In $\S2$ and $\S3$, we briefly describe our data sets, the parameterization of the AGN HOD, and the theoretical modeling of the 2PCF. We present the results of our HOD modeling in $\S4$. Finally, we discuss the implications of our results in $\S5$ and summarize them in $\S6$. Throughout the work we assume a spatially flat, $\Lambda$CDM cosmology: $\Omega_{m}=0.26$, $\Omega_{\Lambda}=0.74$, $\Omega_{b}=0.0435$, $n_{s}=0.96$, $\sigma_{8}=0.78$, and $h=0.71$. We quote all distances in comoving $\hMpc$ and masses in units of $\hMsun$ unless otherwise stated.

\begin{figure*}[t]
\begin{center}
\begin{tabular}{c}
        \resizebox{8cm}{!}{\includegraphics{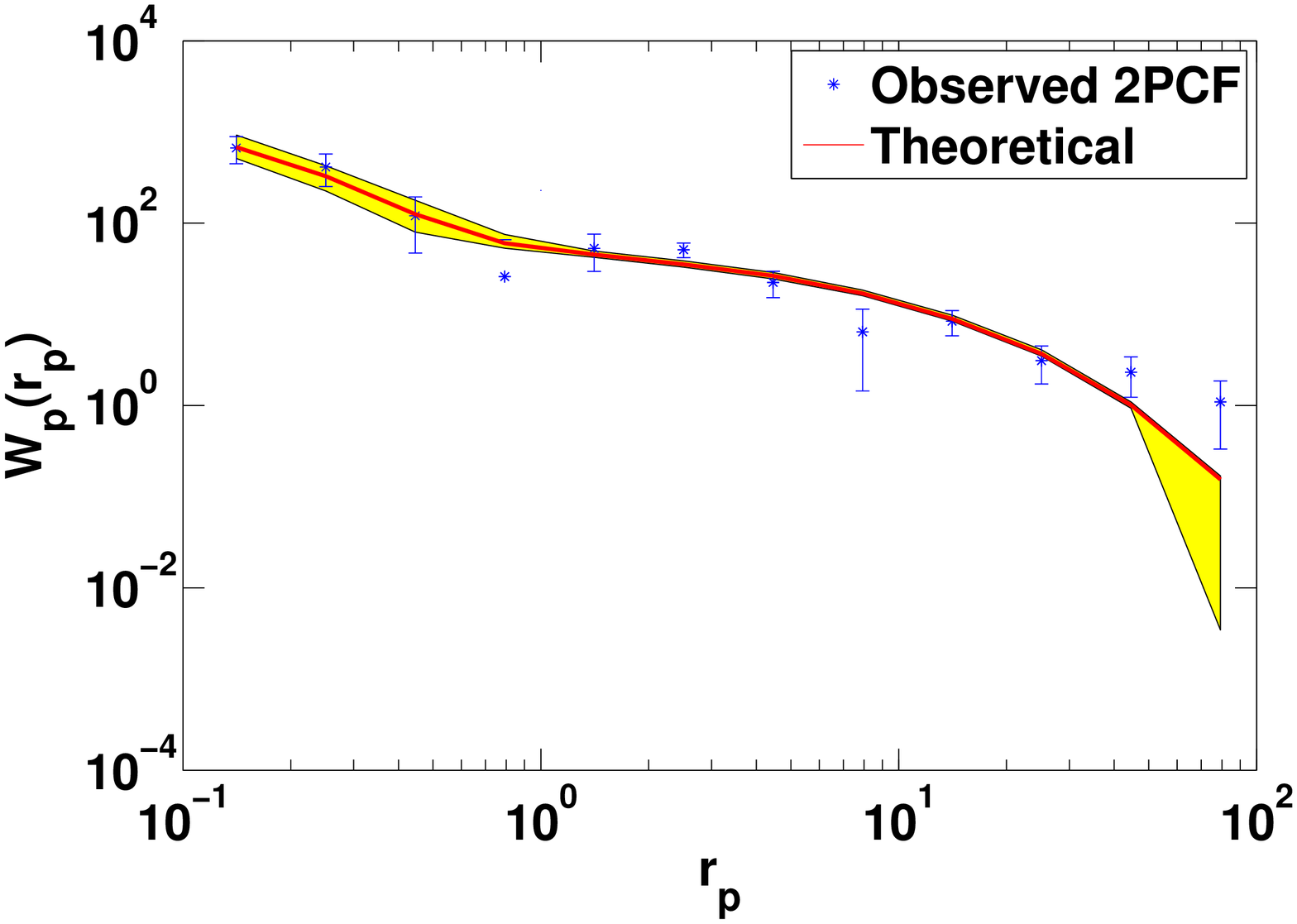}}
        \resizebox{8cm}{!}{\includegraphics{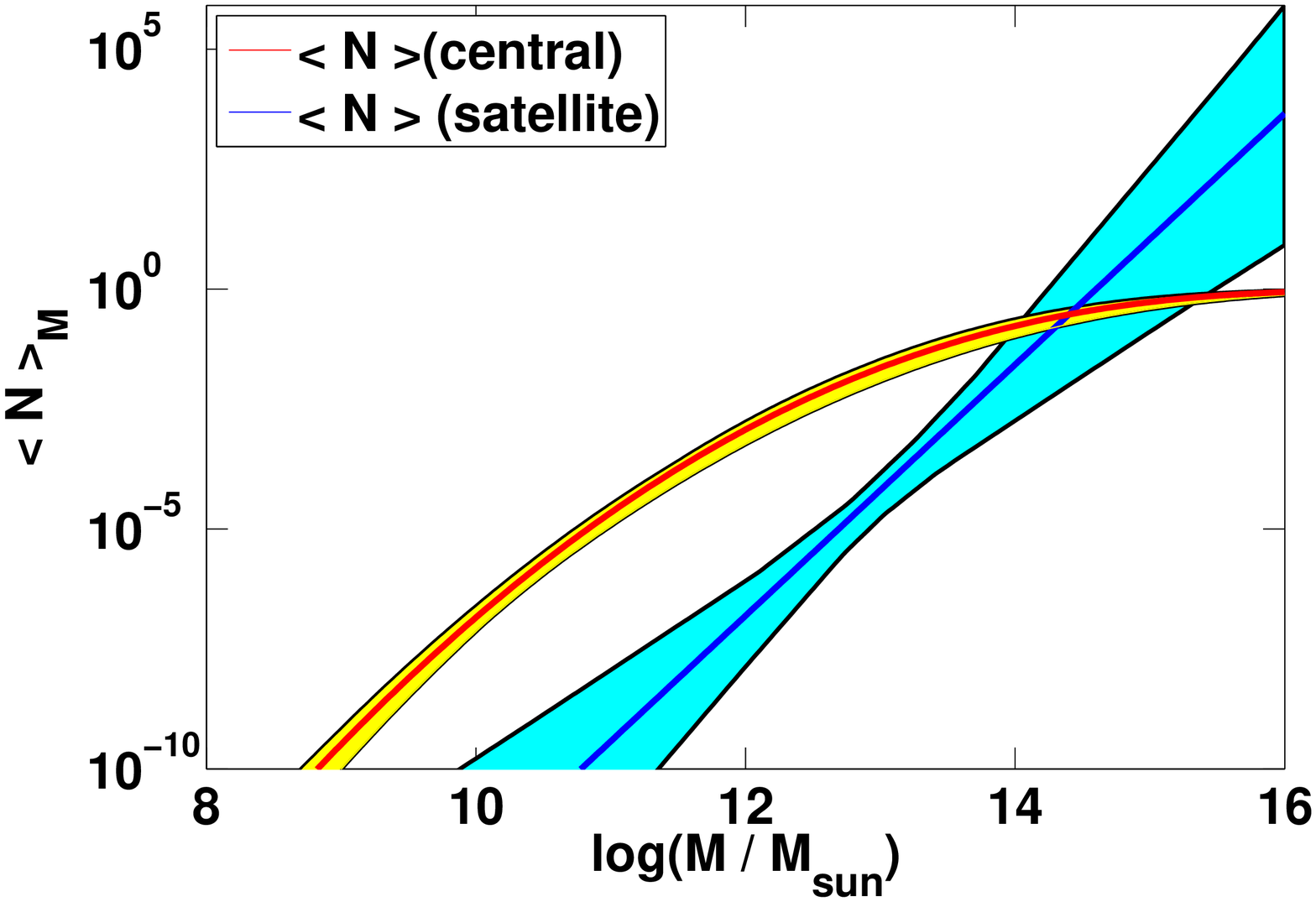}}\\
         \resizebox{8cm}{!}{\includegraphics{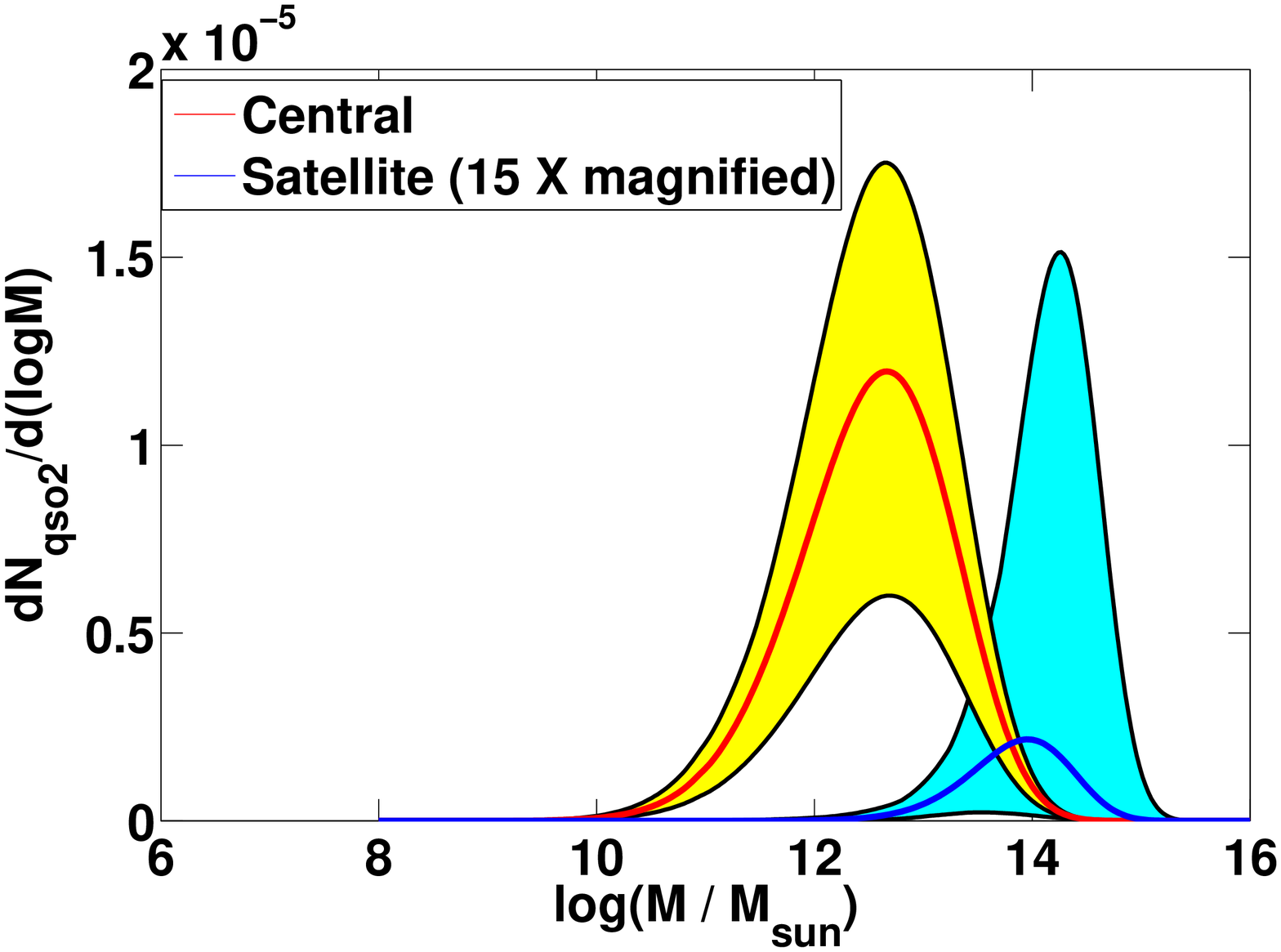}}
         \resizebox{8cm}{!}{\includegraphics{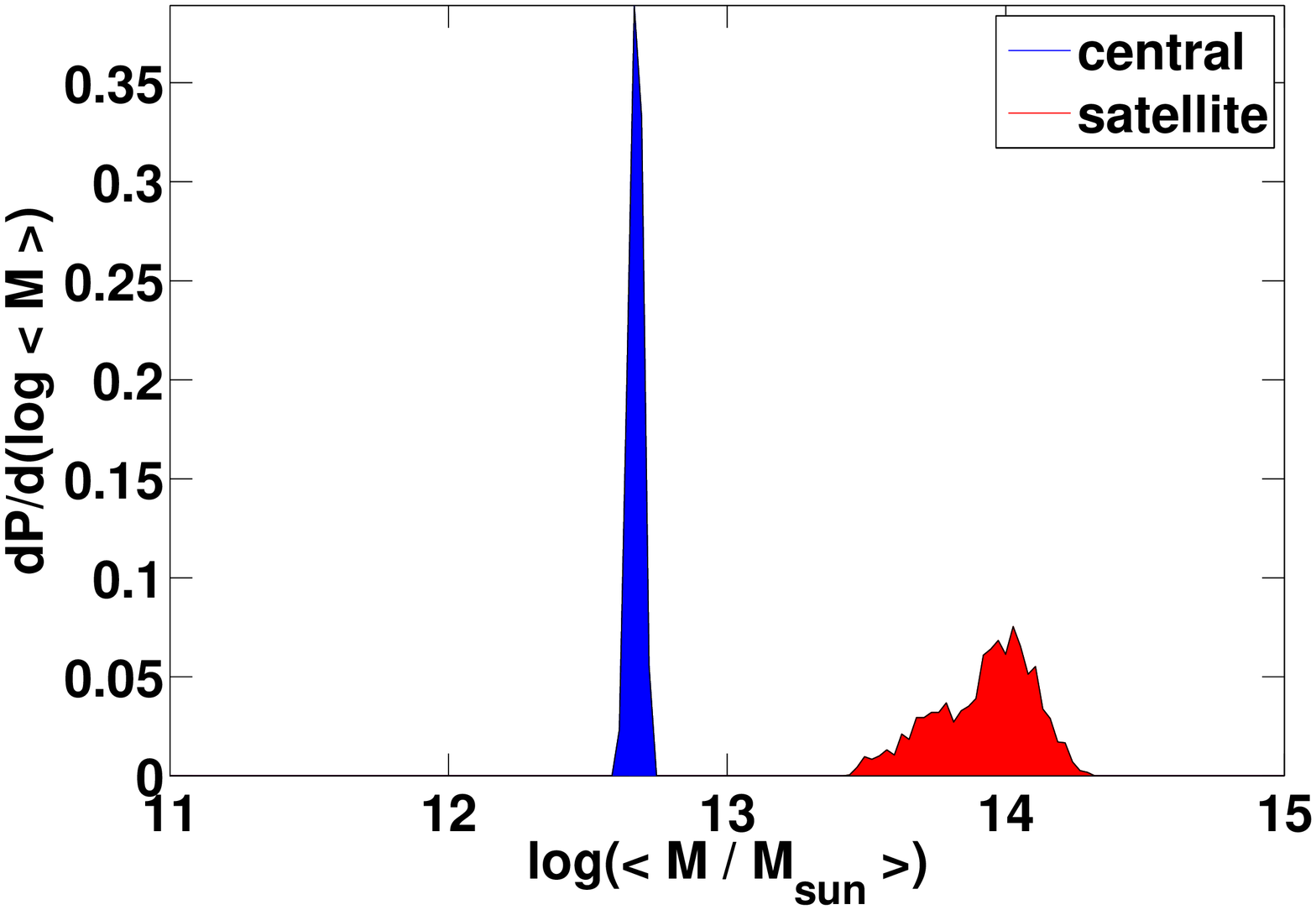}}\\
          \end{tabular}
 \caption{{\bf Top Left} : Projected 2PCF of the WISE-SDSS quasar sample (median redshift 1) as a function of clustering scale. The measurements are taken from \citet{dipompeoetal14}. The red line is the theoretically best fitted 2PCF and the yellow region corresponds to the theoretical error bar. {\bf Top Right} : The Mean Occupation Function \citep{chatterjeeetal12}, constructed from the best-fit parameters, as a function of halo mass. Red line (with yellow error bar) and blue line (with cyan error bar) are the MOFs of the ccentral and satellite quasars respectively. {\bf Bottom Left} : The distribution of central (red) and satellite (blue) quasar abundances in Dark Matter Halos as a function of halo mass. It is constructed by convoluting the MOF of central and satellite quasars with the HMF of \citet{jenkinsetal01}. {\bf Bottom Right} : The normalized distribution of the medians of central (red) and satellite (blue) quasar abundances that reproduce theoretical 2PCF curves consistent with the observed one within a defined $\delta \chi^2$ range as has been discussed below. This provides a probability distribution of finding the peaks of central and satellite quasar abundances in DM halos as a function of halo mass.}
\end{center}
\end{figure*}

\section{Datasets}

The projected $2PCF$ of quasars, that is used in this work, is constructed from the clustering sample of D14. We refer the reader to D14 for a detailed description of the datasets and the measurements. Here we describe the main features of the data. 

The clustering sample is selected from the all-sky catalog of WISE. WISE has mapped the sky in four wavebands at $3.4$, $4.6$, $12$ and $22$ $\mu m$, referred to as $W1$, $W2$, $W3$ and $W4$, with angular resolutions $6.1′′$ , $6.4′′$ , $6.5′′$ , and $12′′$ , respectively. Both obscured (QSO2) and unobscured (QSO1) quasars are observable in $mid-IR$ wavelength range of WISE as the  hot gas in AGN is responsible for an increasing power-law spectrum at longer wavelengths \citep[e.g.,][]{lacyetal04, sternetal05, donleyetal07, lacyetal13}. A simple color cut at $W1-W2 > 0.8$ for objects with $W2 < 15.05$ is used for selecting $249,169$ AGN candidates from the all-sky data in the region $135*<RA<226*$ and $1*<DEC<54*$.

The WISE selected sample is then matched with SDSS-DR8 $r-band$ data using a $2''$ radius and accepting only the closest match. After removing galactic and lunar contaminations by applying different masks, the final sample is a population of $177,709$ WISE selected quasars over an area of $3289$ $ deg^{2}$. The separation of the two types of quasars (type-1 and type-2) is realized by applying the optical-IR color cut at $r-W2 > 6$ \citep[e.g.,][]{hickoxetal07}. The WISE selected quasars having no SDSS counterpart are marked as the  obscured ones resulting to a final sample of $74889 (42 \%)$ obscured and $102740 (58 \% )$ unobscured quasars. For $r$ and $W2$ distributions of these samples and the ($r-W2$) color distribution we refer the reader to Fig.\ 3 and Fig.\ 4, respectively, of D14.

The redshifts of the quasar sample (Fig.\ 5 of Di14), are a mixture of both spectroscopic and photometric measurements. Both the mean and the median redshifts of the unobscured quasars are $z \sim 1.04$ with a standard deviation of $0.58$. Quasars in the sample lie in the redshift range from $z \sim 0.1$ to $2.8$. To compute the number density of quasars we adopt the following technique. Hence the lower limit is set on number density by dividing the total number of unobscured quasars by the net volume of the sphere of comoving radius corresponding to $z \sim 2.8$ and solid angle of $3289$ $ deg^{2}$. The upper limit  is set by assuming all the quasars to be confined within the median redshift, that is $0.1$ to $1.04$, within the given solid angle cone.

\begin{figure*}[t]
\begin{center}
\begin{tabular}{c}
        \resizebox{8cm}{!}{\includegraphics{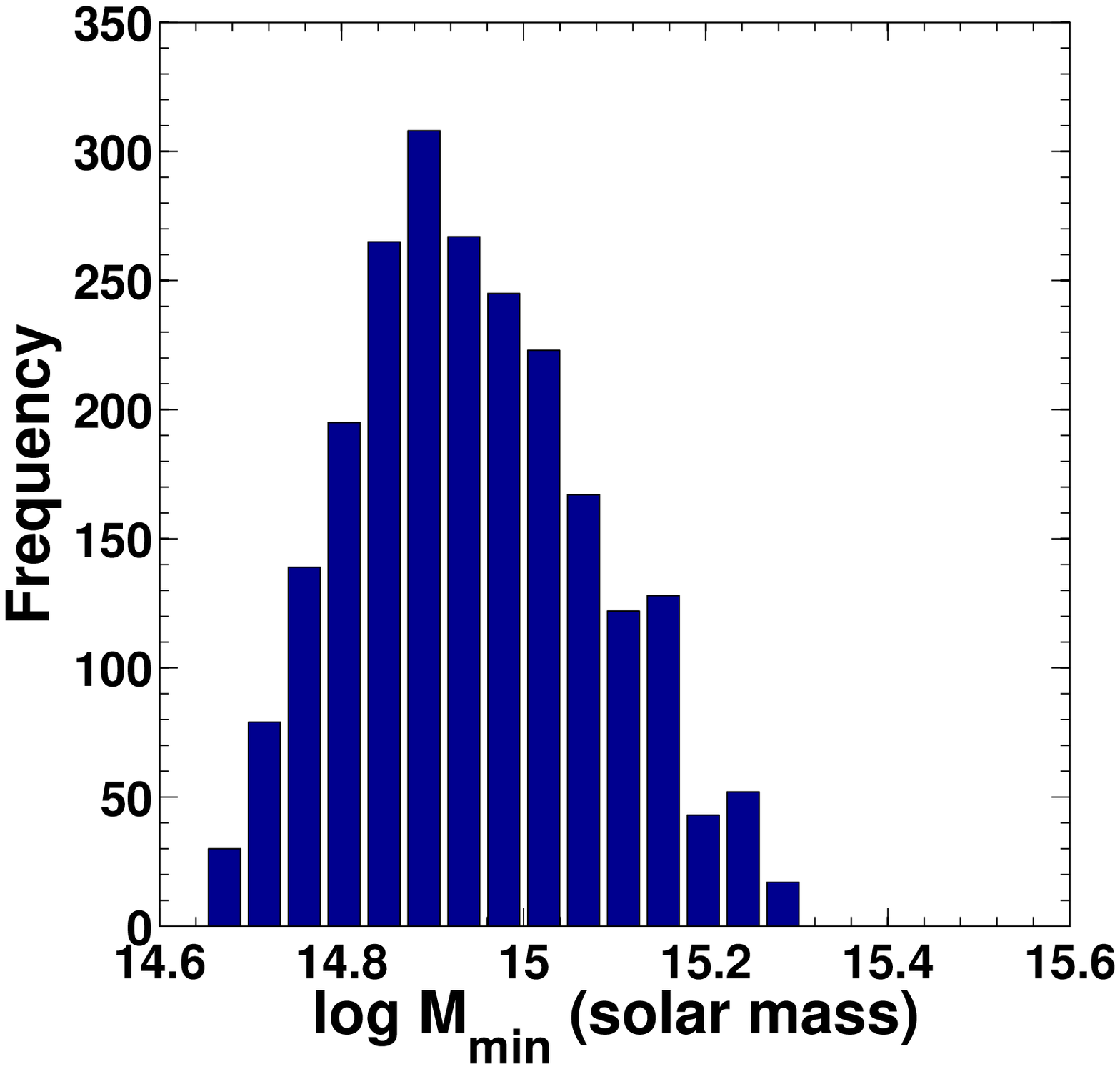}}
        \resizebox{8cm}{!}{\includegraphics{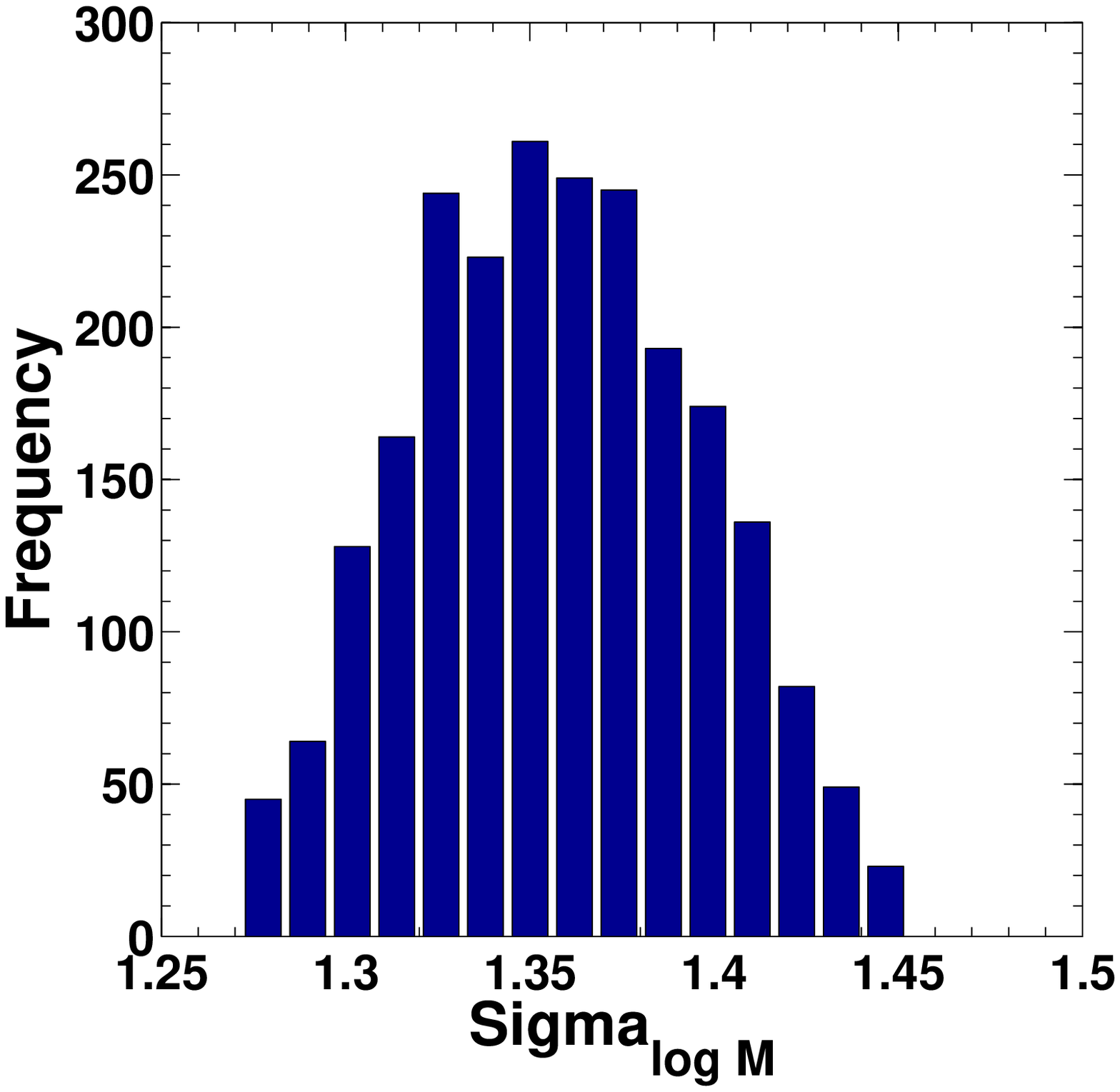}}\\
         \resizebox{8cm}{!}{\includegraphics{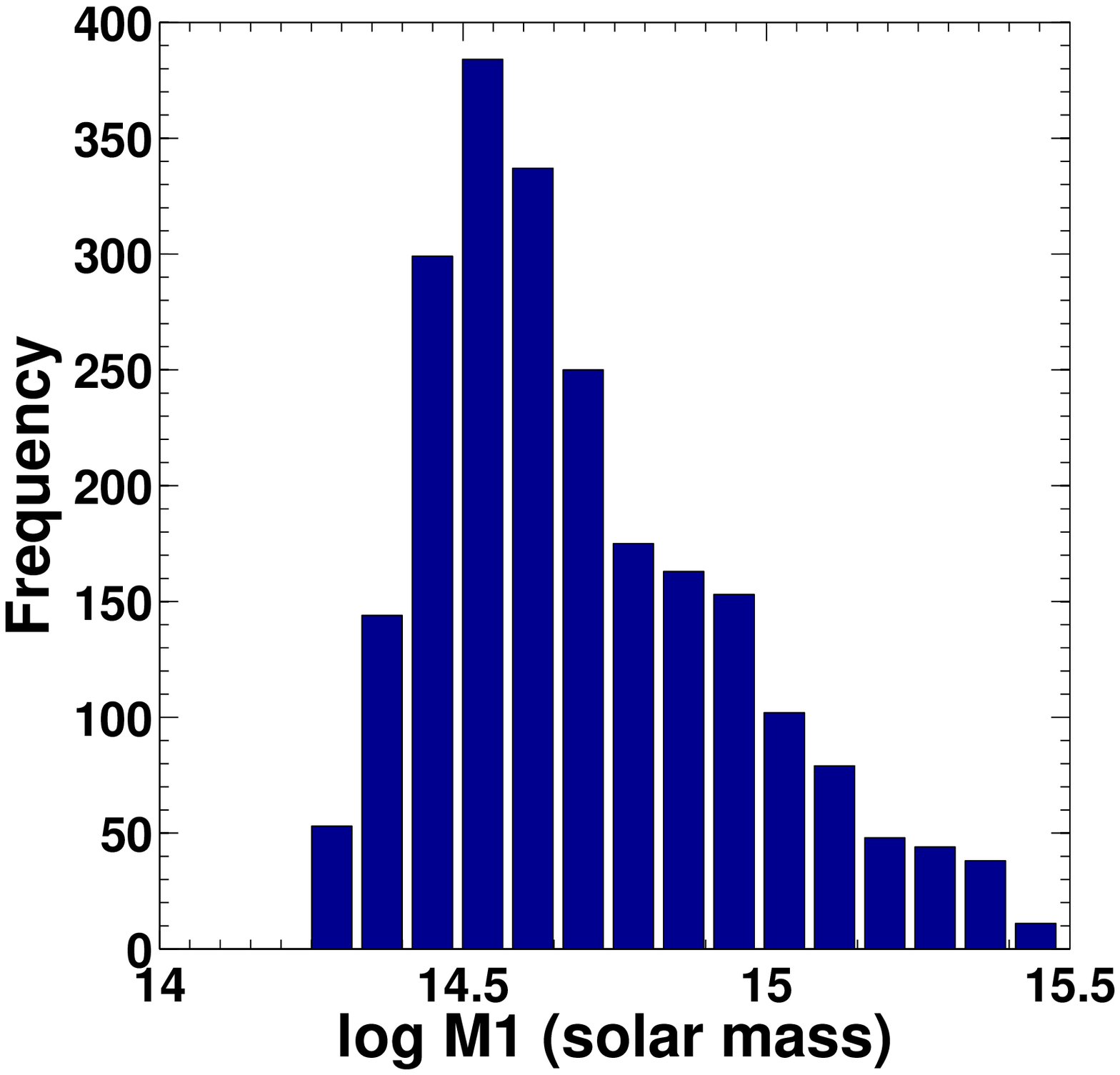}}
         \resizebox{8cm}{!}{\includegraphics{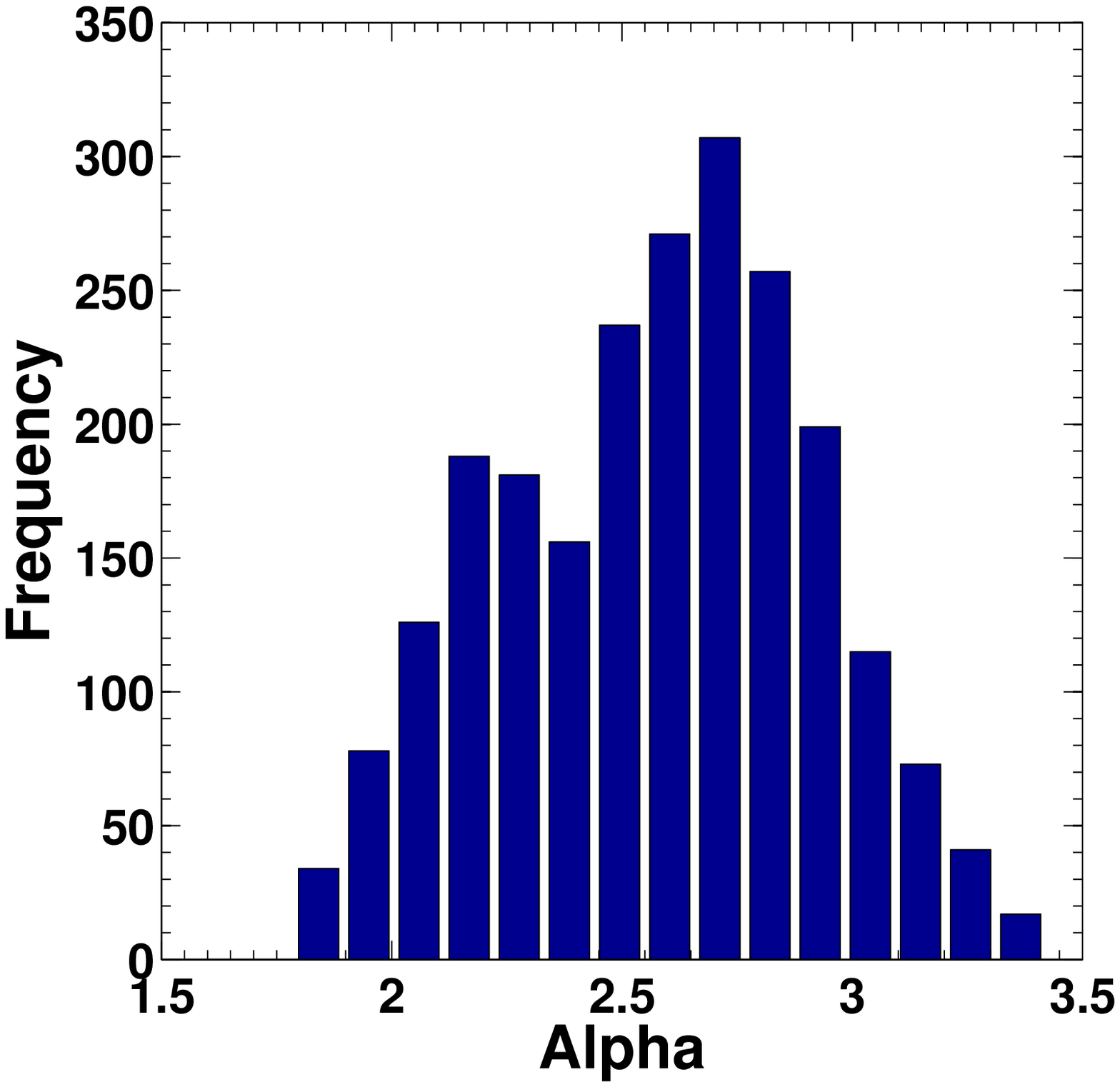}}\\
          \end{tabular}
 \caption{{\bf Top Left} : The histograms of the values of parameters $M_{min}$, $\sigma_{logM}$, $M_1$ and $\alpha$, respectively, as are generated by the MCMC code to fit D14 data. Only those sets of parameters have been considered for which $\chi^2$ lies within $\mu_{chi^2} + \sigma_{\chi^2} = 13.24$. The best-fit values with the required uncertainties are as follows : $M_{min} = 8.5 (^{+11.8} _{-4.0}) \times 10^{14} M_{\odot}$, $\sigma_{logM} = 1.36 (\pm 0.1)$, $M1 = 4.9 (^{+21.0} _{-3.2} ) \times 10^{14} M_{\odot}$ and $\alpha = 2.49 (^{+0.93} _{-0.70})$.}
\end{center}
\end{figure*}

\begin{figure*}[t]
\begin{center}
\begin{tabular}{c}
        \resizebox{8cm}{!}{\includegraphics{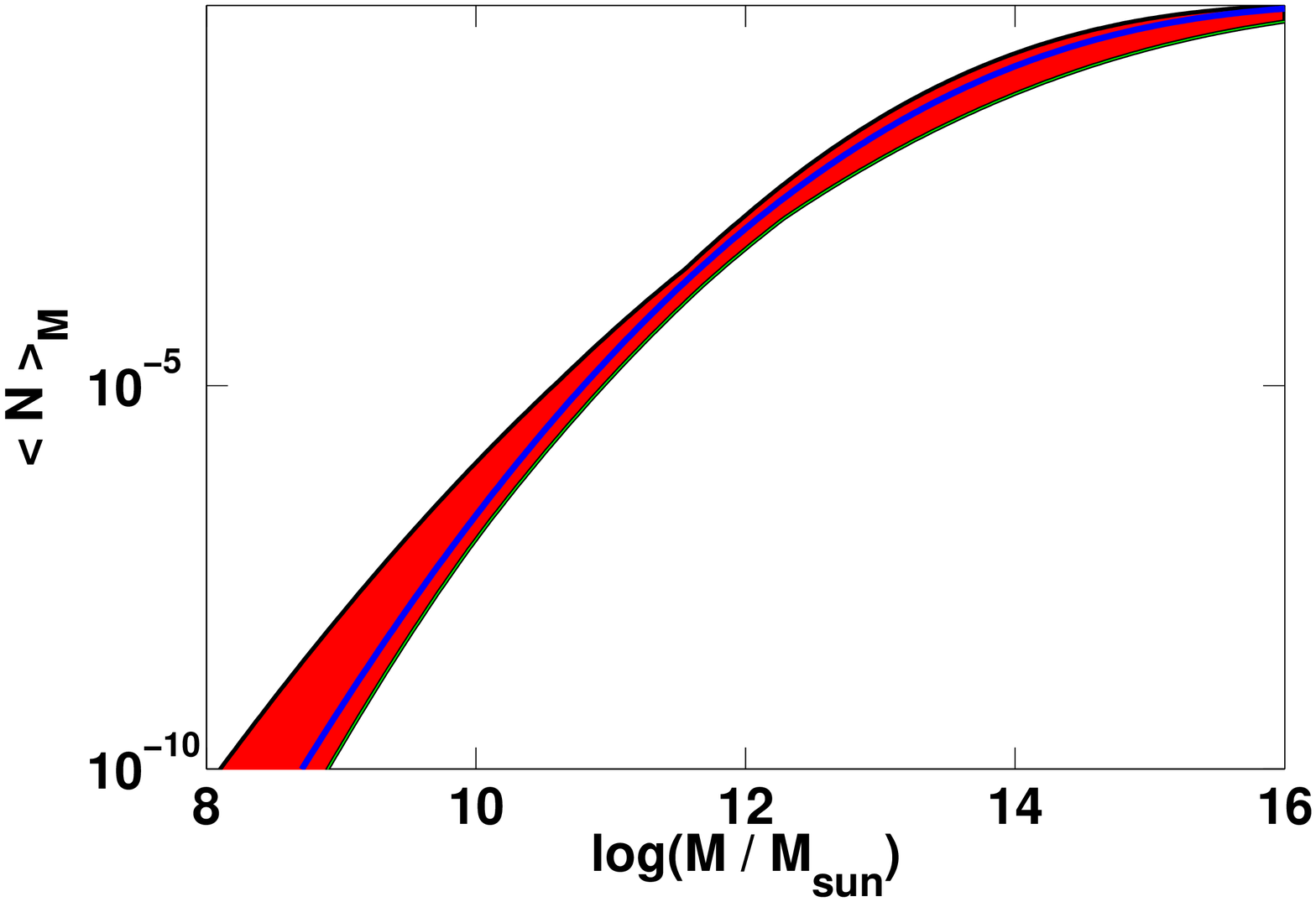}}
        \resizebox{8cm}{!}{\includegraphics{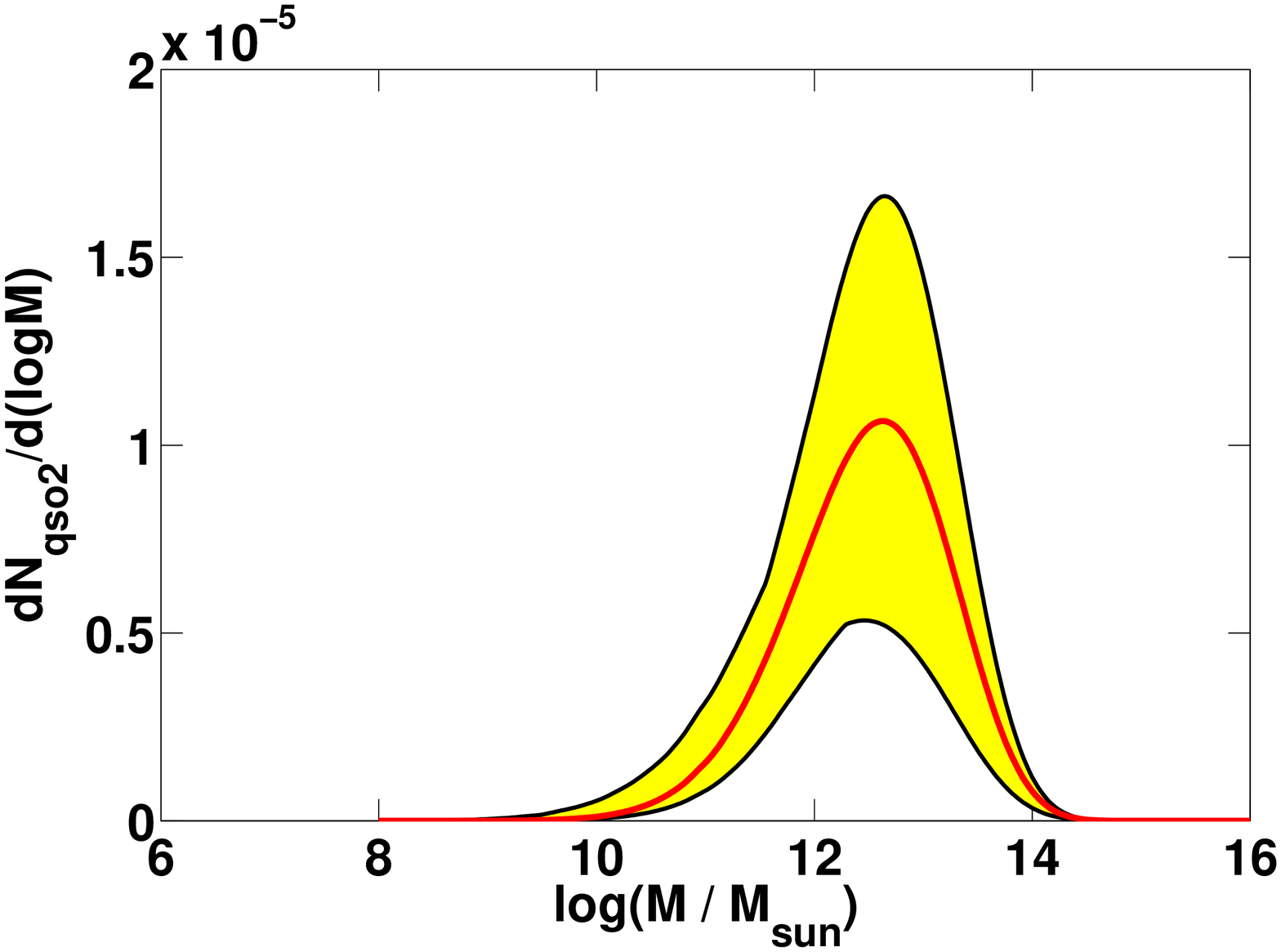}}
          \end{tabular}
 \caption{Left :The Mean Occupation Function of Quasars assuming central-only (2 parameter) model. Right : The distribution of Quasar abundance in dark matter halos as a function of Halo mass.}
\end{center}
\end{figure*}

\begin{figure*}[t]
\begin{center}
\begin{tabular}{c}
        \resizebox{8cm}{!}{\includegraphics{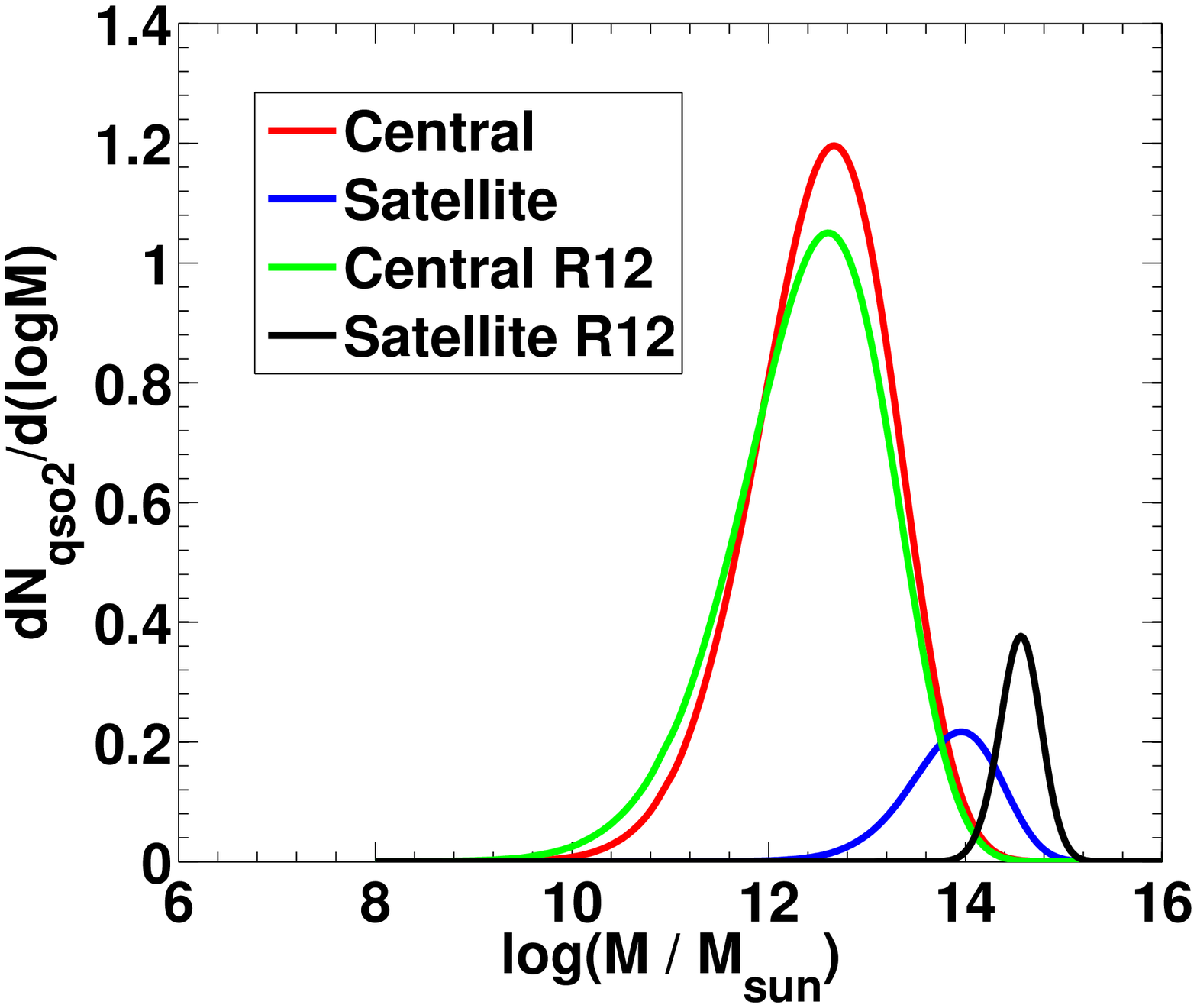}}
          \end{tabular}
 \caption{The comparison between the distribution of WISE-selected SDSS DR8 quasars in dark matter halos as a function of host halo mass and the same for SDSS DR7 quasars as has been found by \citet{richardsonetal12}.}
\end{center}
\end{figure*}

\section{Methodology}

The halo occupation distribution formalism allows us to extract the full distribution of the host dark matter halos of quasars from the 2PCF. For a given cosmological model the typical host masses of quasars can be obtained via bias measurements \citep[e.g.,][]{jing98,shethetal01}. However, those simple bias estimates do not allow us to obtain the full halo distribution of quasars and also do not make any attempt to distinguish between central and satellite quasars. In that sense the HOD provides a more complete description of the connection between quasars and their host halos. In the following sections, we introduce our quasar HOD parameterization and our methodology to model the 2PCF.

\subsection{Halo Occupation Distribution of Quasars}

The HOD of quasars is characterized by $P(N|M)$, the conditional probability that a halo of virial mass $M$ contains $N$ quasars combined with the spatial and velocity distributions of quasars within halos. In principle, $P(N|M)$ could be fully specified by determining all its moments observationally from the quasar clustering at each order. Moreover the HOD can be used to construct any statistic like the void probability distribution, pairwise velocity distribution and others with a known set of cosmological parameters. For our purpose of modeling the 2PCF, we only need the description of the first two moments, $\langle N(M) \rangle$ and $\langle N(N-1) \rangle_{M}$ \citep{b&w02}. The HOD is assumed to be dependent only on the halo mass since the assembly bias effect is assumed to be small for the massive halos that typically host quasars \citep[e.g.,][]{bondetal91}.

The Mean Occupation Function (MOF) or the first moment of the probability distribution $P(N|M)$ is defined as the average number of quasars lying in DM halos as a function of host halo mass. The MOF is taken to be a sum of a softened step function for central and a modified power-law for the satellite fraction of quasars \citep{chatterjeeetal12}, a model that has been developed from cosmological hydrodynamic simulations of AGN growth and feedback. The HOD model is given as, 

\begin{equation} 
{\langle N(M)\rangle}_{cen} = \frac{1}{2}\left[1+{\rm erf}\left(\frac{{\rm log} M-{\rm log} M_{\rm{min}}}{\sigma_{\rm{log M}}}\right)\right], \nonumber
\end{equation}

\begin{equation} 
{\langle N(M)\rangle}_{sat} = \left(\frac{M}{M1}\right)^{\alpha} \exp \left(-\frac{M_{\mathrm{cut}}}{M} \right), \nonumber
\end{equation}

\begin{equation}
\langle N(M)\rangle = {\langle N(M)\rangle}_{cen} + {\langle N(M)\rangle}_{sat}
\end{equation}

where $\langle N(M)\rangle$ is the mean number of quasars lying in halos of mass $M$,  $M_{\rm{min}}$ is the host halo mass at which the average number of quasars per halo is  $0.5$, $\sigma_{\rm{log M}}$ is the transition width of the softened step function, $M1$ gives the reference of the higher mass scale at which the satellite fraction follows a power-law, $\alpha$ is the power-law index, and $M_{cut}$ is the lower mass range at which the satellite fraction falls off exponentially. To do a more constrained fit we excluded $M_{cut}$ and did a 4-parameter modelling of the MOF, with the satellite MOF given as :
\begin{equation} 
{\langle N(M)\rangle}_{sat} = \left(\frac{M}{M1}\right)^{\alpha} \nonumber
\end{equation}

For a given halo mass, satellite quasars in simulations are found to follow an approximate Poisson distribution \citep[e.g.,][]{degrafetal11b, chatterjeeetal12}. Thus for simplicity we assume a Poisson distribution and a nearest integer distribution for the satellite and central quasar occupation numbers respectively. Following \citet{richardsonetal12}, we assume that the halo occupations of central and satellite quasars are uncorrelated with each other. This is in accordance with the studies of \citet{chatterjeeetal12}, where they found no evidence of a correlation between the activity of central and satellite black holes in a hydrodynamic simulation.

To obtain the host dark matter halo population of quasars we convolve the MOF with the halo mass function (HMF). We use the HMF of \citet{jenkinsetal01} in our current model. We model the radial distribution of satellite quasars within halos as an NFW profile \citep{nfw97} with the concentration-mass relation from \citet{bullocketal01},
\begin{equation}
c(M,\,z)= \frac{c_{0}}{1+z} \left( \frac{M}{M_{*}} \right)^{\beta}, 
\end{equation}
where $M_{*}$ is the nonlinear mass for collapse at $z=0$, and $\beta=-0.13$. We use $c_{0}=32$, which happens to be consistent with the high concentration values of locally observed AGN profiles \citep[e.g.,][]{l&m07}. R12 verifies that the modeling is weakly sensitive to the choice of $c_{0}$

\subsection{Calculation of the 2-point Correlation Function}

The quasar 2PCF, $\xi_{q}(r)$, is given as the excess probability of finding quasar pairs separated by a spatial distance $r$ over a random distribution \citep{peebles80}, $P(r) = n^2[1+\xi(r)]dV_1dV_2,$ where $n$ is the number density of quasars in the survey volume. It can be decoupled into contributions from intra-halo pairs, $\xi_{1h}(r)$, and inter-halo pairs, $\xi_{2h}(r)$. The inter-halo or two-halo term is approximated as \citep{b&w02}
\begin{equation}
\xi_{2h}(r) \approx \biggl[n_{q}^{-1} \int_{0}^{\infty} dM \frac{dn}{dM} \langle N(M)\rangle b_{h}(M)\biggr]^{2} \xi_{m}(r),
\end{equation}
where $n_{q}$ is the quasar number density, $dn/dM$ is the differential halo mass function, $b_{h}(M)$ is the halo bias factor, and $\xi_{m}(r)$ is the 2PCF of matter. We identify the bracketed term as the quasar linear bias factor, $b_{q}$. The intra-halo or one-halo term is expressed as
\begin{equation}
1+\xi_{1h}(r) \approx \frac{1}{4 \pi n_{q}^{2} r^{2}} \int_{0}^{\infty} dM \frac{dn}{dM} \left\langle N\left(N-1\right)\right\rangle _{M} \frac{dF_M}{dr},
\end{equation}
where $F_M(r)$ is the average fraction of same-halo pairs at separations $\le r$. The two-halo term depends only on $\langle N(M) \rangle$, while the one-halo term depends on the second moment $\left\langle N\left(N-1\right)\right\rangle _{M}$ and the radial profile of the spatial distribution of quasars through $F_M(r)$.

For calculating the 2PCF, D14 uses the Landy and Szalay estimator, given as \citep{l&s93} 
\begin{equation}
\xi(r) = \frac{(DD_{(r)}-2DR_{(r)}+RR_{(r)})}{RR_{(r)}}, 
\end{equation} 
where $n_R=n_D$. Here $DD_{(r)}$, $RR_{(r)}$ and $DR_{(r)}$ are defined as the number of point pairs separated by $r$ in the observed data, point pairs separated in a random distribution and the number of cross-pairs in the stacked random distribution on the data respectively. From the 3D correlation function we can define the projected 2PCF.

The projected 2PCF is the line-of-sight integral of the $\xi(r)$ \citep{d&p83}.
\begin{equation}
w_p(r_p) = 2 \int_0^{\pi_{max}} \xi(r)d\pi, 
\end{equation}
where $r_{p}$ is the comoving transverse separation and $\pi$ is the line of sight distance such that $r = \sqrt{r_p^2+\pi^2}$. In this work, instead of the correlation function in configuration space, D14 measures the clustering statistic in angular coordinates. The projected angular 2PCF, defined as the excess probability (over a random distribution) of finding quasar pairs separated by an angle $\theta$ on the celestial sphere within a solid angle $d\Omega$, \citep{peebles80} is given by
\begin{equation}
dP = n[1+\omega(\theta)]d\Omega
\end{equation}
For our work we employ an approximate technique to go from angular to spatial projected 2PCF. We discuss our methodology below. 

Let $\theta$ be the angular separation of galaxy pairs, corresponding to a comoving transverse separation $r_{p}$. According to the definition of the angular 2-point correlation function $w_p(\theta)$, the pair count for pairs with separation between $r_p$ and ($r_p+dr_p$) should be
\begin{equation}
N(r_p) = \sigma \times [1+w_p(\theta)] \times 2 \pi r_p dr_p, 
\end{equation}
where $\sigma$ is the surface density of objects (quasars in this case).
We can also calculate the pair count from the $3-D$ correlation function $\xi(\sqrt{r_p^2+\pi^2})$, that is
\begin{equation}
N(r_p) = \int n \times 2 \pi r_p dr_p \times [1+\xi(\sqrt{r_p^2+\pi^2})] d\pi, 
\end{equation}
where $n$ is the number density of galaxies. 

If we consider a periodic cubic box of size $L$, and if the actual number and the surface densities (projected over the full size $L$) of quasars are $n$ and $\sigma$ respectively then we can write $\sigma=n \times L$. If we equate Eq.\ 10 to Eq.\ 11 and consider $L=\pi_{max}$, that is the depth of the survey, we have
\begin{equation}
w_p(\theta)=\int_0^{\pi_{max}} \xi(\sqrt{r_p^2+\pi^2}) d\pi / L
\end{equation}
The projected 2PCF is the line-of-sight integral of $\xi(r)$ \citep{d&p83}.
\begin{equation}
w_p(r_p) = 2 \int_0^{\pi_{max}} \xi(\sqrt{r_p^2 +\pi^2})d\pi
\end{equation}
Hence we can approximately write $w_p(\theta)\times \pi_{max} = w_p(r_p)$,
where $\pi_{max}$ should be understood as the depth of the survey. In our work we used $pi_max = ,$ which is the co-moving distance to $z=1.04$ (median redshift of the sample).


We note that the clustering sample of D14 has been constructed over a wide range of redshift. Hence calculating the 2PCF at the median redshift can be interpreted as an average over the redshift intervals \cite{richardsonetal13}. However, the modeling uses halo properties (e.g., mass function, bias factor) and the redshift evolution of the halo properties are not accounted for in this calculation. \citet{richardsonetal12} have shown that the true HOD can be interpreted as the HOD for objects at the median redshift (within the errors of the measurement), if the 2PCF measured over a wider redshift range is statistically consistent with the actual 2PCF of the same sample at the median redshift. We adopt the above interpretation in this work with the assumption that the clustering evolves weakly with redshift. We refer the reader to R12 and \citet{richardsonetal13} for additional discussion on the limitations of this interpretation.

\section{Results}

To model the 2PCF, we use the routine developed by \citet{zhengetal07}. The code uses the Markov Chain Monte Carlo (MCMC) algorithm in the four-dimensional parameter space. Using the underlying halo mass function from \citet{jenkinsetal01}, the code populates a virtual sky with points following the \citet{chatterjeeetal12} MOF model (Eqns.\ 1 and 2). Following the prescription of \citet{richardsonetal13} we calculate the $\chi^{2}$ value of each point in the parameter space using the diagonal elements of the covariance matrix \citep[see, e.g., the appendixes of][]{myersetal07a,rossetal09}. Each $\chi^{2}$ value accounts for the combined uncertainties of the 2PCF values and the number density of quasars. In our code dark matter halos are defined as objects with a mean density of $200$ times that of the background density \citep[for details about the routine see][]{zhengetal07, richardsonetal12, richardsonetal13}.

The MCMC contains $100,000$ points in the HOD parameter space, and the set of parameters with the minimum $\chi^2$ value plugged back into \citet{chatterjeeetal12} MOF gives the best-fit theoretical model. The error on the best-fit value is computed in the following way. If the degrees of freedom of the $\chi^2$ distribution is $d$ then the theoretical mean of the $\chi^2$ distribution is $\mu_{\chi^2} = d$ and the standard deviation is $\sigma_{\chi^2} = \sqrt{2d}$. So all the points in parameter space having $\chi^2$ in the range $(\mu_{\chi^2} \pm \sigma_{\chi^2}) $ are statistically consistent with the minimum $\chi^2$ point within $1 \sigma$. The envelope to all possible MOF of \citet{chatterjeeetal12} model with these sets of parameters, in the $(\mu_{\chi^2} \pm \sigma_{\chi^2}) $ zone, define the error range in the MOF and hence that propagates to give the error range in the final distribution.


In this work we did two sets of modelling: firstly the 4-parameter fit incorporating both central and satellite quasars in the model and using the complete 2PCF data of D14. The satellite fraction coming out to be negligible, as is physically expected in case of quasars, motivated us to perform a 2-parameter, hence more constrained, fit on the truncated two-halo tail of projected 2PCF data where the halos are populated only with central quasars. We now present our results in the following subsections. In case of the 4 parameter MOF model, the 2PCF ranges from $0.14$ to $79.25$ $h^{-1} Mpc$. While fitting the 2 parameter model the 2PCF points below the typical halo size will have no contribution to the 1-halo term. Hence in the second case the 2PCF was trucated below $0.8$ $h^{-1} Mpc$.

\subsection{The Four-parameter model}

In the top left panel of Fig.\ 1, we show our four-parameter HOD fit of the 2PCF of WISE selected obscured quasars at $z\sim1.04$. With four parameters and twelve data points, combined with the quasar number density, we have nine degrees of freedom. The shaded envelope represents the error (computed according to the method described in \S 3.2) on our best-fit theoretical model. The best fit set of parameters are as follows : $M_{min} = 8.5 (^{+11.8} _{-4.0}) \times 10^{14} M_{\odot}$, $\sigma_{logM} = 1.36 (\pm 0.1)$, $M1 = 4.9 (^{+21.0} _{-3.2} ) \times 10^{14} M_{\odot}$ and $\alpha = 2.49 (^{+0.93} _{-0.70})$. The distribution of the values of the parameters, as generated by the MCMC code is shown in the histograms in Fig.\ 2. The distributions of only those points in parameter spance have been taken which fall in $1 \sigma$ zone around the theoretical $\mu_{\chi^2}$. The histograms crudely do resemble Gaussian forms around the best-fit values, as is expected from the MCMC code. The best-fit set of parameters correspond to the point in the parameter space having minimum $\chi^2 = 11.96$, in the $\chi^2$ space with $9$ degrees of freedom. In the top right panel of Fig.\ 1 we show the MOF from the best-fit HOD model, decomposed into its central (dashed line) and satellite (dot-dashed line) components. The shaded regions refer to the uncertainties in our estimate of the MOF.


In the bottom left panel of Fig.\ 1 we show the host halo mass distribution of quasars. The convolution of MOF with the HMF gives the actual distribution of quasar abundance as a function of host halo mass. The central and satellite distributions of quasars are shown in Fig.\ 1 (with the satellite fraction magnified 15 times). The peak of the satellite fraction is two orders of magnitude lower than that of central fractions, which is expected since the probability of finding two bright quasars in a single DM halo is extremely low. The error-range is plotted using parameters within the $\delta\chi^2$ range. The central population peaks at DM halo of $(5 \pm 1.0) \times 10^{12} M_{\odot}$. The satellite population peaks at $8 (^{+7.8} _{-4.8}) \times 10^{13} M_{\odot}$


In the bottom right panel of Fig.\ 1 we show the distributions of the median halo mass scales of central and satellite quasars which are representatives of the probability distribution of the peak halo mass scale from the MCMC chains. The theoretical mean of the $\chi^2$ distribution is $9$ and standard deviation is $\sigma_{\chi^2} = \sqrt{9 \times 2} = 4.24$. So points with $\chi^2 < (9+4.24)=13.24$ fall within the $1\sigma$ range around the theoretical minimum $\chi^2$, and are statistically consistent with the best-fit set. Hence each of those sets of parameters reproduce a distinct MOF and hence a distinct HOD which is statistically consistent with the theoretical best-fit distribution. The median of all such HODs have been found and Fig.\ 1 shows the histogram of those medians. It provides the uncertainly $\Delta_M$ for the peak of the population distributions.



\subsection {The Two-parameter model}

To do a more constrained fit and to check the robustness of the model we performed the HOD modeling of central-only quasars. For the large scale clustering, the 2-halo term contributes to the 2PCF hence we truncated our 2PCF data below $0.8 h^{-1}Mpc$. The best fit MOF and the host halo distributions are plotted in Fig.\ 3. We note that the host halo distributions obtained from the 2 parameter model is exactly identical to the host halo distribution of central quasars obtained from the 4 parameter model. 

Our results are in excellent agreement with R12. For R12, the distribution of central quasars peaked at a halo mass of $(4.1 \pm 0.4) \times 10^{12} h^{-1}M_{\odot}$ whereas our central distribution peaked at a mass scale of $(5 \pm 1.1) \times 10^{12} h^{-1}M_{\odot}$.

The comparison between the distribution of unobscured quasars from our four-parameter model and the same from \citet{richardsonetal12} has been shown in Fig.\ 4. In R12 the median halo masses of central and satellite quasars lie in the range $M_{\mathrm{cen}} = 4.1^{+0.3}_{-0.4} \times 10^{12} \; h^{-1} \; \mathrm{M_{\sun}}$ and $M_{\mathrm{sat}} = 3.6^{+0.8}_{-1.0}\times 10^{14} \; h^{-1} \; \mathrm{M_{\sun}}$, respectively. The central distribution is in great agreement with our results. There is a significant difference in the satellite distribution and hence in satellite fraction as well, which is $f_{\mathrm{sat}}=(7.4 \pm 1.4) \times 10^{-4}$ in R12 and is $f_{\mathrm{sat}}= 5.5 (^{+35} _{-5.0})\times 10^{-3}$ from our results. Our measured satellite fraction is one order of magnitude higher with an even larger upper bound. The huge uncertainty in the satellite distribution from our work can be attributed to the scarcity of $2PCF$ data points in the one-halo scale.


\section{Discussion of Results and Future Work }

According to the AGN unification theory, the central SMBH and accretion disk of a quasar are surrounded by an optically thick dusty torus (Urry \& Padovani 1995). The obscuration of the central broad line region by the torus due to certain inclination angles of the symmetry axis with the line of sight causes the two distinct population of quasars namely obscured and unobscured types. If the quasar classification is based on orientation theory then one would not expect any statistical difference between the environments of these two classes of quasars.  Neither should there be any statistical difference between the host halos of optically-bright $QSO1$ and IR-selected unobscured quasars. 

In contrast to the orientation theory other authors proposed an evolutionary theory of AGN 
(e.g., Sanders et al, 1987; Hopkins et al.\ 2005; Hopkins et al.\ 2008). Hopkins et al.\ (2006) proposed a merger-driven unification model which says that AGN are triggered by halo mergers and eventual galaxy collisions. Galaxy merger though provides abundant matter for near-Eddington accretion on to the SMBH, it also triggers starburst (Cavaliere \& Vittorini 2000) and enshrouds the region with optically thick dust, hence triggering IR-bright $QSO2$. The stronger correlation of merger and star-formation with $QSO2$, compared to $QSO1$, has been studied by Chen et al.\ (2014). Since they are driven by halo mergers so they are expected to have higher small scale clustering. With the advent of more accretion AGN feedback sets in \citep[e.g.,][]{m&n07}, and drives away the gas and dust around it, (e.g., Somerville et al.\ 2008).

The brightest of these dust obscured quasars, blow away the dust due to feedback flows and becomes an optically bright quasar and enters the unobscured phase (e.g., Hopkins et al.\ 2005). But not all of the initial obscured ones in a single halo are expected to go to the unobscured phase, since feedback from a bright {\it Type-1} will inhibit other {\it Type-1} formation in the same halo \citep[e.g.,]{choietal13}. Even if other {\it Type-1}  do develop that will not be simultaneous; where as the {\it Type-2} formations were more or less simultaneous owing to major halo mergers. Hence there would be a significant loss of satellite population in the transition from {\it Type-2} to {\it Type-1}. Hence the satellite fraction of {\it Type-1} is also expected to be much less. 

In our analysis the similarity in the distributions of the halo mass of the central quasars imply that the large-scale distributions of the two types of quasars (namely optically selected and IR- selected) are identical. This is in accordance with the orientation theory of AGN unification. Now considering IR-bright {\it Type-1} to be an intermediate phase between QSO2 and QSO1, we should also expect higher satellite fraction in the IR sample compared to the optically bright sample. Hence our results, though consistent within the statistical error range, does not naturally follow the predictions of the orientation theory of quasar unification. The evolutionary theory of quasars can however provide some insights towards explaining our results and our results in agreement with what can be expected from the evolutionary theories of AGN evolution. Similar results have been proposed by DiPompeo et al.\ (2014) from simple bias measurements. 

We note that our results need to be explored further to understand whether the higher satellite fraction indicate any link between halo, galaxy and quasar co-evolution or it is a manifestation of systematic effect in our datasets. The median mass of the satellite fraction being lower than R12 might seem to contradict our hypothesis (that few among these IR-bright $QSO1$ satellites finally become optically bright $QSO1$), they might seem to represent two different populations --- but are not! The small satellite peak in R12 is just the higher mass tail of this broader satellite distribution of IR-selected quasars. It is expected that the central $QSO1$ in the smaller halos will suppress other $QSO1$ growth by its inter-galactic feedback, while the satellite IR-bright quasars in bigger and more massive halos are expected to survive and go into optically bright phase. Hence the higher mass tail of the IR-selected satellite quasar population survives to give the higher mass peak in R12 results. 

The strong agreement of the central quasar population is also something that is expected. The $QSO1$ formation and its feedback in one halo has no effect whatsoever on $QSO1$ triggering in other halos. So, though they affect the satellite transition, yet the central triggering in different halos being quite independent --- it keeps the 2-halo clustering statistic unaffected. That also explains the similarity in 2PCF of $QSO1$ and $QSO2$ at scales larger than $1 h^{-1} Mpc$ (DiPompeo et al, 2014). Hence the large-scale clustering of $QSO1$ and $QSO2$ are expected to be identical. This is in complete agreement to our result of having the central distribution of IR-selected quasars matching brilliantly with R12 central distribution.

Our future goal is to expand this work with IR-selected obscured {\it Type-2} quasars and see if the IR-bright QSO1 falls in between the two : $QSO2$ and optically bright $QSO1$ --- which would reinforce our hypothesis. In the galaxy SMBH co-evolutionary theories : X-ray bright quasars play a significant role, hence we would like to compare our results with similar work done on X-ray bright quasars by Richardson et al.\ (2013). We would also like to compare our work, which is a phenomenological and model-dependent technique, with similar work done on measurement of the mean occupation function of quasars through direct observations (Chatterjee et al.\ 2013, Chakrabarty et al.\ 2016 in prep). There are scopes of improvement in our work, and we would like to check the robustness of our model through further analysis and mock-data fittings, to see whether we can put some observational constraints that might in near future break the statistical degeneracies which were unavoidable in our work. 

The goal of this project is to probe the orientation versus the evolutionary theories of quasar unification from the cosmological perspective, and to find the missing links in the picture of galaxy-SMBH co-evolution with the underlying large-scale distribution of dark matter in the universe. Although our results are in accordance with the orientation theory at large, they tend to conflict with some aspects of the orientation based AGN unification model.  We have shown for the first time from a robust halo occupation technique that AGN classification should be revisited in light of the cosmological co-evolution of AGN with galaxies and dark matter halos in the Universe.

\section*{Acknowledgments}

I would like to thank  Dr. Michael DiPompeo (Dartmouth College) for providing me the 2PCF data, Prof. Zheng Zheng (University of Utah) for the MCMC code, Jonathan Richardson (University of Chicago) for providing his results for me to compare with --- and all three of them for helping me out with every doubt I had and everything that I ever asked. I cordially thank Prof. Adam Myers (University of Wyoming) and Prof. Ryan Hickox (Dartmouth College) for their unprecedented help and support. I would like to thank Prof. Ritaban Chatterjee, Dhruba Dutta Chowdhury, Rudrani Kar Chowdhury, Sunip Kumar Mukherjee and all other group members of PresiPACT for their constant help and support regarding everything, from physical understanding of cosmology and astrophysics to overcoming computational challenges, and more importantly for providing me the necessary motivation. I would like to thank Prof. Kanan Kumar Datta for agreeing to be the co-reader of my project report. Above all, the person who made me dream about cosmology and made me reach it as far as I have, who provided me such a wonderful project, inspired and motivated me to work through it --- my teacher, mentor and guide Prof. Suchetana Chatterjee, I can never thank her enough.

\bibliography{mybib}{}

\end{document}